\documentclass[11pt,epsf]{article}
\usepackage{appb,epsfig}
\begin{document}

\def\la{\;
\raise0.3ex\hbox{$<$\kern-0.75em\raise-1.1ex\hbox{$\sim$}}\; }
\def\ga{\;
\raise0.3ex\hbox{$>$\kern-0.75em\raise-1.1ex\hbox{$\sim$}}\; }
\def\rl{\;
\raise0.3ex\hbox{$\rightarrow$\kern-1.0em\raise-1.1ex\hbox{$\leftarrow$}}\;}

\title{ NEUTRINO EMISSION DUE TO ELECTRON BREMSSTRAHLUNG
        IN SUPERFLUID 
NEUTRON-STAR CORES}
\author{A.D.~KAMINKER
\address{
        A.F.~Ioffe Physical Technical Institute,
        194021 St.Petersburg, Russia   \\
        E-mail: {\tt kam@astro.ioffe.rssi.ru}
        }
\and P.~HAENSEL
\address{
       N.~Copernicus Astronomical Center,
       Bartycka 18, 00-716 Warszawa, Poland   \\
       E-mail: {\tt haensel@camk.edu.pl}
       }
}
\headtitle{Neutrino emission due to electron bremsstrahlung}
\headauthor{A.D.~Kaminker, P.~Haensel}
\maketitle
\vskip 6mm
\centerline{published in: {\it Acta Physica Polonica B }{\bf 30} (1999) 
 1125}
\vskip 4mm
\begin{abstract}
We study 
neutrino energy emission
rates
(emissivities) due to
electron bremsstrahlung produced by $ee$ and $ep$ collisions
in the superfluid neutron star cores.
The neutrino emission due to $ee$ collisions is
shown to be the dominant neutrino reaction
at not too high temperatures ($T \la 10^8$ K)
in dense matter if
all other
neutrino reactions involving
nucleons are 
strongly  suppressed
by  neutron and proton superfluidity.
Simple practical expressions
for the $ee$ and $ep$ neutrino emissivities
are obtained.
The efficiency of
various neutrino reactions in the superfluid neutron--star
cores is discussed
for the cases of standard neutrino energy losses
and the losses enhanced by the direct Urca process.

\end{abstract}

\section{Introduction}
In the first $10^5$ years of their life, neutron stars (NSs)
cool mainly via 
neutrino emission
from their cores.
This emission results from weak interaction
processes involving baryons and leptons.
It is important, thus, to study various neutrino
reactions. For simplicity, we will mainly consider
the NS cores composed of neutrons $n$, with some
admixture of protons $p$ and electrons $e$.
The neutrino reactions 
in the neutron-star  cores
are
traditionally divided into two groups,
which differ drastically by their efficiency and
produce the {\em standard}
or {\em rapid} cooling (e.g., ref.\ [1]).

The standard cooling is mainly provided
by the neutron and proton branches of the modified
Urca reactions [2, 3]
\begin{equation}
       n+N \rightarrow p+N+e+\bar{\nu}_e~, ~~~~~
       p+e+N \rightarrow n+N+\nu_e~,
\label{Mur}
\end{equation}
where $N$ is a nucleon ($n$ or $p$),
and by the nucleon--nucleon ($nn$, $np$, and $pp$) bremsstrahlung processes
\begin{equation}
       N+N \rightarrow N+N+\nu+ \bar{\nu}~.
\label{NN}
\end{equation}
The rapid cooling
is strongly enhanced by the direct
Urca
processes 
\begin{equation}
       n \rightarrow p+e+\bar{\nu}_e~, ~~~~~
       p+e \rightarrow n+\nu_e~,
\label{Dur}
\end{equation}
which can operate [4]
only in the central regions of rather massive NSs
with proton fraction exceeding 11\%
of the total baryon density.
If the direct Urca
reaction is allowed (in a 
non-superfluid
 NS),
the neutrino emissivity is typically
4--5 orders of magnitude higher than in the standard
reactions (\ref{Mur}) and (\ref{NN}).

It is widely accepted (see, e.g.,\ [5, 6]
and references therein),
that neutrons and protons in a NS core can be in a superfluid state.
The superfluidity is generally thought to be of BCS
type, produced by
attractive component of nuclear forces.
Various microscopic theories
predict very different critical
temperatures of neutron and proton superfluids,
$T_{{\rm c}n}$ and $T_{{\rm c}p}$, which
range from $10^8$ to $10^{10}$~K.
The neutron superfluidity in the NS cores is
generally believed to be produced
by triplet-state pairing of neutrons
(with anisotropic superfluid energy gap),
while the proton superfluidity is due to the singlet-state
pairing (with isotropic gap).
If $T \ll T_{{\rm c}n}$ and $T \ll T_{{\rm c}p}$
the superfluidity strongly
suppresses all the neutrino reactions
(\ref{Mur})--(\ref{Dur})
involving nucleons
(e.g.,\ [1, 3, 7]).

Onset of the neutron or 
proton superfluidity
with the decrease of temperature
in a cooling NS initiates also an additional
specific neutrino reaction which can be considered as
neutrino production due to Cooper pairing of quasinucleons
$\widetilde{N}$
\begin{equation}
          \widetilde{N} + \widetilde{N} \to \nu + \bar{\nu}~.
\label{CP}
\end{equation}
The emission due to singlet-state pairing of neutrons
has been calculated by Flowers et al.\ [8] and
Voskresensky \& Senatorov [9, 10]  while the emission
due to triplet-state pairing of neutrons and singlet-state
pairing of protons has been determined by Yakovlev et al.\ [11, 12]
and Kaminker et al.\ [13].
The effect is much stronger for neutrons than for protons
due to the smallness of axial--vector electroweak currents
for protons.

In contrast to the non-superfluid cores, where the main
neutrino emission is produced either by the modified
or by the direct Urca processes
(depending on the equation of state and density),
very different neutrino mechanisms can dominate
in the superfluid 
cores, depending on the
relative
magnitudes
of 
$T$, $T_{{\rm c}n}$ and $T_{{\rm c}p}$.
In particular,
Cooper-pair neutrino emission (\ref{CP})
can be very important if
$0.1 \, T_{{\rm c}N} \la  T \la T_{{\rm c}N}$
(e.g., ref.\ [12]).
At lower temperatures, $T \ll T_{{\rm c}N}$, this emission
falls exponentially.

On the other hand, as mentioned by Schaab et al.\ [14],
an additional neutrino emission mechanism,
the neutrino-pair bremsstrahlung due to $ep$ scattering
\begin{equation}
         e + p  \to   e + p + \nu + \bar{\nu}~,
\label{ep}
\end{equation}
can be important in the superfluid NS cores.
This mechanism, and the closely related mechanism of
neutrino bremsstrahlung due to $ee$ scattering
\begin{equation}
         e + e  \to   e + e + \nu + \bar{\nu}
\label{ee}
\end{equation}
has been considered briefly by Kaminker et al.\ [15].
The latter authors have estimated the neutrino energy
emission rates (emissivities) in these reactions
using simplified criteria of similarity of neutrino
emission and
electron conduction processes.
It is quite evident that these reactions, especially
the neutrino $ee$ bremsstrahlung, can be
important in highly superfluid NS cores,
where all the reactions involving nucleons are drastically
suppressed by the nucleon superfluidity. 
Thus,  rigorous calculation
of the emissivities in the reactions (\ref{ep}) and (\ref{ee})
deserves special study.

Notice that the
neutrino bremsstrahlung due to $ee$ scattering
of nondegenerate electrons
at not too high densities, $\rho \la 10^{10}$ g cm$^{-3}$
(corresponding to the outer NS crust),
was considered by Cazzola \& Saggion [16, 17].
The authors obtained general but cumbersome
expressions for the
emissivity [16]
and performed numerical calculations
using Monte Carlo method [17].
The main result of their consideration was that
the emissivity due to the $ee$  bremsstrahlung  did not exceed $ 1\% $
of the total emissivity 
from  other neutrino
reactions
in a NS crust (pair annihilation, photoneutrino
and plasmon decay). We will show that this is not so in
a NS core.

The aim of the present article is
to carry out an accurate calculation of the neutrino emissivities
for the bremsstrahlung reactions (\ref{ep}) and
(\ref{ee}),
and to analyze the efficiency of
various neutrino production mechanisms (\ref{Mur})--(\ref{ee})
in  superfluid NS cores.

\newpage
%
\section{Superfluid 
neutron-star cores}
Mass density 
of the matter 
in a NS
core is expected to range from about 0.5$\rho_0$
at its outer edge [18],
to (5--10)$\rho_0$ at the stellar center,
where $\rho_0 = 2.8 \times 10^{14}$ g cm$^{-3}$
is the saturation
density of nuclear matter.
At $\rho \la 2 \rho_0$, 
neutron-star matter
consists  mostly of neutrons, 
with a few percent admixture of protons,
electrons, and possibly muons.
The fraction of muons is usually much smaller
than that of electrons, and a simple $npe$  model
is a good approximation.
At higher densities,
$\rho \ga 2 \rho_0$, other particles may appear, such as
hyperons, condensed pions or kaons, or even free
(deconfined) quarks.
Neutrino emissivity
of dense neutron-star matter
 with different compositions has been
reviewed by Pethick [1].
For simplicity, we will neglect possible presence of
hyperons and exotic particles, and consider
$npe$ matter only.
All constituents of the matter
are strongly degenerate.
The electrons
form an ultrarelativistic and almost ideal gas, with
the electron Fermi-momentum
\begin{equation}
          p_{{\rm F}e} = \hbar (3 \pi^2 n_e)^{1/3} = 648.6~
          ( n_e / n_0 )^{1/3} \, m_e c~,
\label{ne}
\end{equation}
where $n_e$ and $m_e$ are  electron number density and
mass, respectively,
and $n_0=0.16$ fm$^{-3}$ is the standard nuclear 
saturation density. The
electron Fermi energy (chemical potential) is
$\mu_e \approx  p_{{\rm F}e} c  \approx$
(100--300) ~MeV.
Neutrons and protons are nonrelativistic
and form strongly interacting Fermi liquids.
The condition of beta-equilibrium implies that the chemical
potentials of particles
satisfy the
equality $\mu_n = \mu_p + \mu_e$.
Charge neutrality leads to the equality of
proton and electron number densities
($n_p=n_e$) and Fermi momenta ($p_{{\rm F}e}=p_{{\rm F}p} $).
The Fermi energy of neutrons is close to that
of electrons, while the Fermi energy of protons
is much smaller (typically, 3--30 MeV).

Both nucleon species, $n$ and $p$, can be in a superfluid
state (Sect.\ 1). Microscopically, the superfluidity
leads to the appearance of the energy gap $\Delta$
in the nucleon dispersion relation.
For a triplet-state neutron superfluid, the gap is
anisotropic along the neutron Fermi surface.
For a singlet-state pairing of protons the gap is isotropic.
The proton gap $\Delta_p=\Delta_p(T)$ depends
only on temperature $T$. In the frame of the BCS theory
(which we adopt further on),
the proton critical temperature $T_{{\rm c}p}$ is related
to the zero-temperature energy gap $\Delta_p(0)$
by $T_{{\rm c}p} = \Delta_p(0)/(1.764 \, k_{\rm B})$
[19],
where $k_{\rm B}$ is the Boltzmann constant.

Superfluidity of neutrons and protons suppresses the main
neutrino generation mechanisms  (\ref{Mur})--(\ref{Dur})
in the NS core and opens an additional
neutrino reaction (\ref{CP}).
Strong superfluidity of neutrons and protons
($T_{{\rm c}n} \gg T$ and $T_{{\rm c}p} \gg T$)
suppresses drastically all neutrino reactions
(\ref{Mur})--(\ref{ep}) involving nucleons.
Superfluidity of protons affects even the
neutrino $ee$ bremsstrahlung (\ref{ee}),
(not very strongly),
by changing screening of $ee$
interaction.

The plasma screening momentum
$q_s$ in $npe$ matter
is given by (e.g., ref.\ [20])
\begin{equation}
    y_s^2 \equiv { q_s^2 \over 4 p_{{\rm F}e}^2 }
            = { e^2 \over \pi \hbar c }
    \left(1 +
    { m_p^\ast ~ p_{{\rm F}p} \over m_e^\ast \, p_{{\rm F}e}}
    D_p \right)~
\label{y_s}
\end{equation}
where $m_e^\ast = \mu_e / c^2$
and $m_p^\ast$ is an effective
proton mass (which determines the proton density of states
near the Fermi level and
which can differ from
the bare mass $m_p$ due to polarization effects in dense matter).
The first term in brackets comes from
the electron screening,
while the second term
is due to the proton
screening. 
The latter term contains
 the function $D_p$.
For nonsuperfluid protons ($T \geq T_{{\rm c}p}$)
one has $D_p=1$ , while for $T < T_{{\rm c}p}$
the factor
$D_p$
describes superfluid reduction of proton screening.
Gnedin \& Yakovlev [20] fitted numerical values of
this factor, calculated for singlet-state proton pairing, by
\begin{eqnarray}
   D_p & = & \left(0.9443 + \sqrt{(0.0557)^2 + (0.1886 y)^2}
              \right)^{1/2}
\nonumber    \\
             &  \times  &
             \exp \left(1.753 - \sqrt{(1.753)^2+ y^2}\right)~,
\label{D_p}
\end{eqnarray}
where $y=\Delta_p(T)/(k_{\rm B}T)$.
According to Levenfish \& Yakovlev [21]
the parameter $y$ at any
$T<T_{{\rm c}p}$ is given by the fitting expression
\begin{equation}
   y \, = \,  \sqrt{1 - { T \over T_{{\rm c}p} } }~
        \left(1.456 - 0.157 ~ \sqrt{T_{{\rm c}p} \over T}  +
         1.764 \, {T_{{\rm c}p} \over T}  \right)~.
\label{gap}
\end{equation}
%
%
\section{General formalism of neutrino $ep$ and $ee$ bremsstrahlung}
Consider two neutrino bremsstrahlung mechanisms
(\ref{ep}) and (\ref{ee})
for stron-gly
degenerate, relativistic electrons
and strongly degenerate, nonrelativistic protons.
We will treat these processes using the standard
perturbation theory of weak interactions.
We will use the units in which
$c=\hbar=k_{\rm B}=1$ and return to ordinary
physical units in the final expressions.

The emissivities $Q_{ep}$ and $Q_{ee}$
[erg cm$^{-3}$ s$^{-1}$] of the processes (\ref{ep}) and (\ref{ee})
can be written as
\begin{eqnarray}
      Q_{12}  & = & { \zeta_{12} \over (2 \pi)^{14}}  \!
              \int \! {\rm d} {\bf p}_1 \! \int \! {\rm d} {\bf p}_2
              \! \int \! {\rm d} {\bf p}'_1
              \! \int \! {\rm d} {\bf p}'_2
              \! \int \! {\rm d}  {\bf k}_\nu \!
              \int \! {\rm d} {\bf k}'_\nu
              \nonumber \\
              & \times &
              \delta^{(4)}(P_1 + P_2 - P'_1 - P'_2 - K) \,
              \omega \, f_1 f_2 (1 - f'_1) (1 - f'_2) \, W,
\label{Qgeneral} 
\end{eqnarray}
where
$P_1 = (\varepsilon_1, {\bf p}_1)$ and 
$P_2= (\varepsilon_2,{\bf p}_2)$ are
4-momenta of two charged particles before scattering,
$P'_1 = (\varepsilon'_1, {\bf p}'_1)$ and
$P'_2= (\varepsilon'_2,{\bf p}'_2)$ are
4-momenta of the same particles
after scattering.
Index 1 labels
an electron and index 2
labels
either a proton or a second electron;
$\zeta_{ep}=1$,
$\zeta_{ee}=1/2$
(to avoid double counting of the same
$ee$ scattering events).
The quantities $K_\nu 
= (\omega_\nu, {\bf k}_\nu)$ 
and $K'_\nu =(\omega'_\nu, {\bf k}'_\nu)$
are 4-momenta of neutrino and
antineutrino, respectively, and
$K = K_\nu + K'_\nu = (\omega, {\bf k})$ is 4-momentum
of the neutrino pair $(\omega = \omega_\nu + \omega'_\nu$ and
${\bf k} = {\bf k}_\nu + {\bf k}'_\nu)$.
Furthermore,
\begin{equation}
       f_i = \left[ 1 + \exp \left( \frac{\varepsilon_i - \mu_i}{T}
                                \right) \right]^{-1}
\label{FDf}
\end{equation}
is a Fermi-Dirac distribution of a particle before scattering
($i$=1, 2), and
$f'_i$ is 
the corresponding  distribution  for a particle
 after scattering.
The $\delta$-function describes
4-momentum  conservation.
Finally,
$W$ is the differential transition rate
\begin{equation}
      W= {G_{\rm F}^2 \over 2} \; {  \sum_{\rm \sigma,\nu} | M_{12} |^2
          \over
         (2 \omega_\nu)(2 \omega'_\nu)
         (2 \varepsilon_1)(2 \varepsilon_2)
         (2 \varepsilon'_1)(2 \varepsilon'_2)}~ ,
\label{W}
\end{equation}
where $G_{\rm F}= 1.436 \times 10^{-49}$ erg cm$^3$
is the Fermi weak interaction constant,
and
$|M_{12}|^2$ is the squared transition amplitude.
Summation is over spin states of
charged particles
before and after scattering
$\sigma=(\sigma_1, \sigma_2, \sigma'_1, \sigma'_2)$
and over the neutrino flavors
$\nu = (\nu_{e}, \nu_{\mu}, \nu_{\tau})$.
The neutrino energies are assumed to be much lower
than the intermediate boson mass
($\sim$80 GeV).
The processes in question are represented by four ($ep$ scattering )
or eight ($ee$ scattering) Feynman diagrams.
Each diagram includes a neutrino-pair emission
four-tail vertex. The amplitudes are
\begin{equation}
M_{ep}=M_A,  \,\,\,  M_{ee}=M_A - M_B~,
\label{M}
\end{equation}
where
\begin{eqnarray}
      M_A  & = &  4 \pi e^2  l^\alpha  \,
            \left[  \frac{1}{t_{22} - q_s^2} \,
            \left( \bar{u}'_2 \gamma^\beta u_2 \right)
            \left( \bar{u}'_1  L^{(1)}_{\alpha \beta} u_1 \right)\right.
\nonumber \\
                      &   +  &
             \left.    \frac{1}{t_{11} - q_s^2}  \,
              \left( \bar{u}'_1 \gamma^\beta u_1 \right)
              \left( \bar{u}'_2 L^{(2)}_{\alpha \beta} u_2 \right)
              \right]~,
\label{M1}   \\
          M_B                & = &
          4 \pi e^2  l^\alpha \,
               \left[   \frac{1}{t_{21} - q_s^2} \,
                \left( \bar{u}'_1 \gamma^\beta u_2 \right)
                \left( \bar{u}'_2 L^{(3)}_{\alpha \beta} u_1 \right)
                \right.
\nonumber   \\
                           &    +    &
                \left.        \frac{1}{t_{12} - q_s^2} \,
                \left( \bar{u}'_2 \gamma^\beta u_1 \right)
                \left( \bar{u}'_1 L^{(4)}_{\alpha \beta} u_2 \right)
                \right],
\label{M2}    \\
                \, & \, & \,
\nonumber      \\
    L^{(1)}_{\alpha \beta} & =  & \Gamma_{e \alpha} \,  G_e(P'_1 + K)
               \,  \gamma_\beta  \,
              + \, \gamma_\beta  \, G_e(P_1 -K) \, \Gamma_{e \alpha}~,
\nonumber    \\
    L^{(2)}_{\alpha \beta} & = & \Gamma_{2 \alpha} \, G_2(P'_2 + K) \,
                  \gamma_\beta  \,
              + \,   \gamma_\beta \, G_2(P_2 - K) \, \Gamma_{2 \alpha}~,
\nonumber  \\
    L^{(3)}_{\alpha \beta} & = & \Gamma_{e \alpha} \, G_e(P'_2 + K)
               \,   \gamma_\beta  \,
              + \, \gamma_\beta \, G_e(P_1 -K) \, \Gamma_{e \alpha}~,
\nonumber    \\
    L^{(4)}_{\alpha \beta} & = & \Gamma_{e \alpha} \, G_e(P'_1 +K) \,
                    \gamma_\beta  \,
              +  \,  \gamma_\beta \, G_e(P_2 -K) \, \Gamma_{e \alpha}~,
\label{OL} \\
              \Gamma_e^{\alpha} & = & C_{eV} \gamma^\alpha +
              C_{eA} \gamma^\alpha \gamma^5~,~~~
              \Gamma_p^{\alpha}  =  C_{pV} \gamma^\alpha +
              C_{pA} \gamma^\alpha \gamma^5~,
\label{Gamma}
\end{eqnarray}
$l^\alpha = (\bar{\psi}_\nu \gamma^\alpha (1 + \gamma^5) \psi_\nu)$
is neutrino 4-current;
$u_1$, $u_1'$, $u_2$ and $u_2'$
denote standard bispinor amplitudes of charged particles;
$\psi_\nu$ and $\psi_\nu'$
are the standard bispinor amplitudes 
of emitted  neutrinos; 
upper bar denotes a Dirac conjugate,
Greek indices $\alpha$ and $\beta$
run over (0,1,2,3),
$\gamma^\alpha$ and $\gamma^5$ are Dirac
matrices;
$G_e(P)$ and $G_p(P)$ are the electron and proton propagators
$G_i(P) = (P_\alpha \gamma^\alpha + m_i)/(P^2 - m_i^2)$.
We use the bare proton  mass $m_2 = m_p$
in the proton propagator and in the normalization
of the proton bispinor amplitudes
$\bar{u}_p u_p = 2 m_p$  in
Eqs.\ (\ref{M1}) and (\ref{OL}).
It will be shown that the final results are independent of
$m_p$ which is a consequence of nonrelativistic approximation
for protons. Nevertheless the results do depend on the
proton density of states near the Fermi level, i.e., on $m_p^\ast$.
An additional term $M_B$ in $M_{ee}$
appears due to identity of colliding electrons; it
contains four diagrams with interchanged final electron states.
In Eqs.\ (\ref{M1})  and (\ref{M2}) we
have used photon propagators in the Feynman gauge
[22] and four familiar kinematic invariants
\begin{eqnarray}
      t_{11}  & = & (P_1 - P'_1)^2 , \,\,\,\,
      t_{22}   =    (P_2 - P'_2)^2 ,
\nonumber   \\      
      t_{21}  & = & (P_2 - P'_1)^2, \,\,\,\,
      t_{12}   =    (P_1 - P'_2)^2,
\label{t}
\end{eqnarray}
where $t_{11}$ and $t_{12}$ are the standard kinematic invariants
denoted usually as $t$ and $u$.
We have added the squared screening momentum
$q_s^2$ (see Sect.\ 2)
to the kinematic invariants in the denominators
of Eqs.\ (\ref{M1}) and (\ref{M2}) to
account for the plasma screening effects.
Furthermore, in Eq.\ (\ref{Gamma})
$C_{iV}$ and $C_{iA}$ are, respectively,
the vector and the axial vector
weak coupling constants
associated with four-tail neutrino-emission vertices
on electron ($i=e$) or proton ($i=p$) lines of the
Feyman diagrams.
For emission of $\nu_e \bar{\nu}_e$ pairs at $i=e$
(charged + neutral currents), one has
$C_{eV} = 2 \sin^2 \theta_{\rm W} +0.5$ and $C_{eA}= 0.5$,
while for emission of $\nu_\mu \bar{\nu}_\mu$ and
$\nu_\tau \bar{\nu}_\tau$
(neutral currents only),
$C'_{eV} = 2 \sin^2 \theta_{\rm W} - 0.5$ and $C'_{eA} = -0.5$.
In this case,
$\theta_{\rm W}$ is the Weinberg angle,
$\sin^2 \theta_{\rm W} \simeq 0.23$.
Neutrino emission
for $i=p$
goes only via neutral currents
and we have
$C_{pV}=(1- 4 \sin^2 \theta_{\rm W})/2 \approx 0.04$ and
$C_{pA}=1.26/2=0.63$, independently of neutrino flavors.

In our case,
neutrino-pair momentum is sufficiently small
($k \sim T \ll p_{{\rm F}e}$), so that we can
use the approximation of Festa \& Ruderman [23] for the electron
propagators:
\begin{equation}
        G_e(P' + K) = \frac{P'_\alpha \gamma^\alpha}
                        {2 (P' K)}~, \, \, \, \, \, \, \,
        G_e(P - K ) = - \frac{P_\alpha \gamma^\alpha}{2(P K)}
\label{Gel}
\end{equation}
and the nonrelativistic approximation for the proton propagators
\begin{equation}
       G_{\rm p}(P'_2 + K) = - G_{\rm p}(P_2 - K) =
        \frac{1+ \gamma^0}{2 \omega}~.
\label{Gpr}
\end{equation}
The standard tedious calculations
using the Lenard identity [22]
\begin{eqnarray}
           &  \,  &
           \int \! {\rm d} {\bf k}_\nu \int \! {\rm d} {\bf k}'_\nu \,
           \delta^{(4)}(K - K_\nu - K'_\nu) \,
           \frac{ K_\nu^{\alpha} {K'}_\nu^{\beta}}{\omega_\nu \omega'_\nu}
           \nonumber  \\
           &  \, &  \, = \, \frac{\pi}{ 6} (K^2 g^{\alpha \beta} +
           2 K^\alpha K^\beta)~,
\label{Identity}
\end{eqnarray}
yield
\begin{eqnarray}
        Q_{12}  & = &
            \frac{ e^4 \, G_{\rm F}^2 }{12 (2 \pi)^{11}} \!
            \, \, \zeta_{12} \!
            \int  {\rm d} {\bf p}_1   \int {\rm d} {\bf p}_2
            \int  {\rm d} {\bf p}'_1  \int {\rm d} {\bf p}'_2 \,
            \frac{\omega}
            {\varepsilon_1 \varepsilon'_1 \varepsilon_2 \varepsilon'_2}
\nonumber   \\
               &  \times  &
             J_{12} \, f_1 f_2 (1 - f'_1) (1 - f'_2)~.
\label{QJ}
\end{eqnarray}
For $ep$ scattering we have
\begin{equation}
         J_{ep} =   \frac{2^6 C_{e+}^2 \, m_p^2 }{(t_{22}-q_s^2)^2}
                  \, \frac{K^2 (P_1 P'_1)
                      (2 \varepsilon_1 \varepsilon'_1 - P_1P'_1)}
                      {(P_1 K)(P'_1 K)}~,
\label{Jpr}
\end{equation}
where $C_{e+}^2=\sum_\nu (C_{eV}^2+C_{eA}^2)$.
Notice that $Q_{ep}$ determined by
Eqs.\ (\ref{QJ}) and (\ref{Jpr}) appears to be very similar
to the emissivity $Q_{\rm br}$ given by
Eqs.\ (A7) and (A13) in the article by
Haensel et al.\ [24] who considered neutrino
bremsstrahlung due to Coulomb scattering of electrons
by atomic nuclei in the NS crust. Equation
(\ref{Jpr}) describes only
neutrino emission associated
with electron lines of the Feynman diagrams
and does not include
analogous emission associated
with proton lines
(the second term in Eq.~(\ref{M1}))
which vanishes in
the limit of nonrelativistic protons.

In the case of $ee$ scattering we have
\begin{equation}
          J_{ee}  = C_{e+}^2 \, K^2 (J_1 + J_2 - J_3)~,
\label{Jee}
\end{equation}
where
\begin{eqnarray}
        J_1 & = & X_1 - I_1~, \quad J_2 = X_2 - I_2~, \quad
          J_3 = Y_1 + Y_2 + Z_1 + Z_2~,
\nonumber   \\
        I_1 & = &
           \frac{2^6}{(t_{11} - q_s^2)(t_{22} - q_s^2)}
           \left[ (P_1 P_2)(P'_1 P'_2)
          + { C_{e-}^2 \over C_{e+}^2} (P_1 P'_2)(P_2 P'_1) \right]
\nonumber  \\
       & \times &
         \left[ \frac{(P'_1 P'_2)}{(P'_1 K)(P'_2 K)} +
         \frac{(P_1 P_2)}{(P_1 K)(P_2 K)} -
         \frac{(P_1 P'_2)}{(P_1 K)(P'_2 K)} -
         \frac{(P_2 P'_1)}{(P_2 K)(P'_1 K)}  \right],
\nonumber    \\
       X_1 & = &  2^6 \, \left[ (P_1 P_2)(P'_1 P'_2) +
         (P_2 P'_1)(P_1 P'_2) \right]
\nonumber    \\
       &  \times &
         \left[\frac{1}{(t_{22} - q_s^2)^2} \; \frac{(P_1 P'_1)}
         {(P_1 K)(P'_1 K)} \,  +  \,
         \frac{1}{(t_{11} - q_s^2)^2} \;  \frac{(P_2 P'_2)}
         {(P_2 K)(P'_2 K)} \right] ,
\nonumber   \\
       Y_1 & = & {2^6 (P_1 P_2) (P'_1 P'_2) \over
                 (t_{11} - q_s^2)(t_{12} - q_s^2)}
\nonumber    \\
     &  \times & \left[ \frac{(P'_1 P'_2)}{(P'_1 K)(P'_2 K)}
         - \frac{(P_2 P'_1)}{(P_2 K)(P'_1 K)} -
         \frac{(P_2 P'_2)}{(P_2 K)(P'_2 K)} \right],
\nonumber     \\ 
       Z_1 & = & { 2^6 (P_1 P_2) (P'_1 P'_2) \over
           (t_{22} - q_s^2)(t_{12} - q_s^2)}
\nonumber    \\
       & \times & \left[ \frac{(P_1 P_2)}{(P_1 K)(P_2 K)}
        - \frac{(P_1 P'_1)}{(P_1 K)(P'_1 K)} -
        \frac{(P_2 P'_1)}{(P_2 K)(P'_1 K)} \right],
\label{JXYZ}
\end{eqnarray}
where $C_{e-}^2=\sum_\nu (C_{eV}^2-C_{eA}^2)$. The expressions
for $I_2$ and $X_2$ are obtained from those for $I_1$ and
$X_1$ by replacing $1 \rl 2$. The expressions for $Y_2$ and
$Z_2$ are obtained from those for $Y_1$ and $Z_1$ by
replacing $1 \rl 2$ and $1' \rl 2'$ 
simultaneously.
The quantities defined in Eqs.\ (\ref{JXYZ}) possess
numerous symmetry properties. For instance, $X_1$
is invariant
with respect to replacing $1 \rl 2$ and $1' \rl 2'$, etc.
The integrand in  (\ref{QJ})  can be shown to 
be highly symmetric as well. Using
the symmetry properties
one can easily prove that
in Eq.\ (\ref{QJ})
it is sufficient to use the expression
\begin{equation}
            J_{ee}   =  2 \,C_{e+}^2
            \, K^2 (X - I_1 - Y_1 - Z_1)~.
\label{Jsym}
\end{equation}
In this case
\begin{equation}
            X   =   2^7 \;
            { (P_1 P_2)(P'_1 P'_2)  +
            (P_2 P'_1)(P_1 P'_2) \over
             (t_{22} - q_s^2)^2} \;
            \frac{(P_1 P'_1)}
            {(P_1 K)(P'_1 K)}  
\label{X}
\end{equation}
corresponds
to ``direct" $ee$ scattering, i.e.,
to square of any single term (amplitude)
in $M_A$ or $M_B$ in Eqs.~(\ref{M})--(\ref{M2});
$I_1$ corresponds
to interference
of amplitudes with exchanging electrons 1 and 2
$(1 \rl 2, 1' \rl 2')$;
while $Y_1$ and $Z_1$ correspond to interference
of amplitudes with mutually
transposed electron final states $1' \rl 2'$.

For further analysis, it is convenient to introduce
an appropriate 4-momentum transfer 
$Q = P'_2 - P_2 = P_1 - P'_1 - K = (\Omega, {\bf q}) $
from particle 1 to particle 2
due to Coulomb interaction.
Using the 4-momentum conserving delta-function in
Eq.~(\ref{Qgeneral})
we can remove the integration over ${\bf p}'_1$ and over $\omega$,
and then replace
the integration over ${\bf p}'_2$ by the integration over ${\bf q}$.
It is convenient also to express $Q_{ep}$ and $Q_{ee}$
in the form suggested by Haensel et al.\ [24]
for the emissivity $Q_{\rm br}$
of the neutrino bremsstrahlung due to scattering of electrons
off atomic nuclei in the NS envelopes:
\begin{equation}
             Q_{12} = \frac{8 \pi G_{\rm F}^2 e^4 C_{e+}^2}
             {567} \, T^6  n_2 L_{12}~.
\label{QL}  
\end{equation}
In a Coulomb liquid of atomic nuclei,
a dimensionless quantity
$L_{12}$ has meaning of a Coulomb
logarithm. For neutrino $ee$ and $ep$ bremsstrahlung,
we are interested in,
$L_{12}$ is a
dimensionless function to be evaluated.
General expressions
for $L_{12}$ are
\begin{eqnarray}
        L_{ep}  & = &
        \frac{567 }{(2 \pi)^{10} T^6 p_{{\rm F}p}^3 }
        \int  {\rm d} {\bf p}_1   \int {\rm d} {\bf p}_2 
        \int  {\rm d} {\bf q}  \int {\rm d} {\bf k} \,\,
         \frac{\omega}{\varepsilon_1 \varepsilon'_1 }  \,
\nonumber   \\
              & \times &  
             \frac{K^2}{(q^2 + q_s^2)^2}
              \frac{(P_1 P'_1)(2 \varepsilon_1 \varepsilon'_1 - P_1 P'_1)}
              {(P_1 K)(P'_1 K)}
              \, f_1 f_2 (1 - f'_1) (1 - f'_2)~,
\label{Lep} \\
             \, & \, & \,
\nonumber    \\
        L_{ee}  & = &
        \frac{567 }{2^{16} \pi^{10} T^6 p_{{\rm F}e}^3 }
        \int  {\rm d} {\bf p}_1   \int {\rm d} {\bf p}_2 
        \int  {\rm d} {\bf q}  \int {\rm d} {\bf k} \,\,\,
        \frac{\omega}{ \varepsilon_1 \varepsilon'_1
        \varepsilon_2 \varepsilon'_2 }
\nonumber   \\
              & \times &  
                   K^2 \, ( X - I_1 - Y_1 - Z_1 )
                  \, f_1 f_2 \, (1 - f'_1) (1 - f'_2)~.
\label{Lee}
\end{eqnarray}
To obtain Eq.\ (\ref{Lep})
we set $\varepsilon_2\varepsilon'_2 = m_p^2$ in the
denominator of (\ref{QJ}) because
the protons are nonrelativistic.

Since charged particles are strongly
degenerate, the main contribution
to $L_{12}$ comes from those transitions, in which
the particle momenta before and after collisions
lie in the narrow
thermal shells around their Fermi surfaces,
$|\varepsilon - \mu| \la T$.
In Eqs.\ (\ref{Lep}) and (\ref{Lee})
we may set
d${\bf p}_i = m_i^\ast p_{{\rm F}i}$
d$\varepsilon_i$  d$\Omega_i$,
where d$\Omega_i$ is solid angle element
in the direction of ${\bf p}_i$.
We also put $\varepsilon_1=p_1$   
in Eqs.\ (\ref{Lep}) and (\ref{Lee}),
and $\varepsilon_2=p_2$
in Eq.~(\ref{Lee}), respectively,
since electrons are ultrarelativistic.

Following Haensel et al.\ [24] we introduce the quantities
${\bf q} = {\bf q}_t + {\bf q}_r  \, \, $,  
${\bf k} = {\bf k}_t + {\bf k}_r \, \, $
and  ${\bf p}''_1 = {\bf p}_1 - {\bf q}_t =
{\bf p}'_1 + {\bf q}_r + {\bf k} \, \, $,
where   ${\bf q}_t$ corresponds to purely elastic Coulomb scattering
while   ${\bf q}_r$   takes into account inelasticity;
vector  ${\bf p}''_1$ is directed along ${\bf q}_r $
but has the same length as ${\bf p}_1$, i.e., $\, |{\bf p}''_1| =
|{\bf p}_1|$;   ${\bf k}_r$ and ${\bf k}_t$ are the orthogonal
vector components parallel and perpendicular to ${\bf p}''_1$,
respectively.
Notice that $q_t = 2 p_1 \sin (\vartheta/2)$,
where $\vartheta$ is an angle
between ${\bf p}_1$  and  ${\bf p}''_1$.
Strong electron degeneracy implies $q_r \ll p_{{\rm F}e} \, $,
$k  \ll  p_{{\rm F}e}$ and $\Omega \ll p_{{\rm F}e}$,
although $q_t$ can be comparable to $p_{{\rm F}e}$ for large-angle
electron scattering events.
From geometrical consideration and energy-momentum conservation
in analogy with expressions given by Haensel et al.\ [24]
we obtain
\begin{eqnarray}
        && q^2 = ({\bf q}_t + {\bf q}_r )^2 =
           q_t^2  +  q_r^2  - q_t^2 \, ( q_r / p_{{\rm F}e})~,
\nonumber \\
        &&    P_1 P'_1  \approx
                     ( q_t^2 + k_t^2  + 2 {\bf q}_t \cdot {\bf k})/2~ ,
\nonumber   \\
        &&    P_1 K   \approx  p_1 (q_r - \Omega) - {\bf q}_t \cdot {\bf k},
                               \, \, \, \,
              P'_1 K  \approx  p_1 (q_r - \Omega)~,
\nonumber   \\
       &&   \Omega + \omega  =  \varepsilon_1 -  \varepsilon'_1
                       \, \,   \approx  \, \,  q_r + k_r~  ,
\nonumber   \\
       &&    K^2  =  \omega^2 - k_r^2 - k_t^2  \, \, \approx \, \,
                     k_0^2 - k_t^2~.
\label{OK}
\end{eqnarray}
Here, $k_0^2 = (q_r - \Omega)(2\omega - q_r + \Omega)$, and the
condition $K^2 > 0$ yields $k_0 \geq k_t$  and   
$q_r \geq \Omega$,  $ \omega \geq (q_r - \Omega)/2 \geq 0 $.

In what follows, we consider
$ep$  and $ee$ scattering separately.

\section{Neutrino $ep$ bremsstrahlung}
Let us evaluate $Q_{ep}$ from
Eqs.~(\ref{QL}) and (\ref{Lep}):
\begin{eqnarray}
          L_{ep} & = &   \frac{567 m_p^\ast }
                {(2 \pi)^{10} T^6 }
                \int {\rm d} \varepsilon_1
                \int {\rm d} \varepsilon_2
                \int  {\rm d} \Omega_1
               \int {\rm d} \Omega_2
               \int_0^{2\pi} {\rm d} \varphi_q
               \int_0^{2 p_{{\rm F}e}} {\rm d} q_t \, q_t
\nonumber    \\               
                \, & \, & \,        
\nonumber     \\
                 \,  & \, &  
               \int_{-\infty}^{\infty} {\rm d} q_r
               \int_{(q_r - \Omega)/2}^{\infty} {\rm d} \omega
               \int_0^{2\pi} {\rm d} \varphi
                \int_0^{k_0} {\rm d} k_t \, k_t \,
                f_1 \,f_2 \,(1- f'_1) \, (1 - f'_2)
\nonumber   \\
                    \, & \, & \,
\nonumber    \\              
                       & \times  &
                {\omega  \over  \varepsilon_1 \varepsilon'_1} \,\,
                \frac{ (k_0^2 - k_t^2) }
              { ( q_t^2 + q_r^2  - \Omega^2 + q_s^2 )^2} \, \,
                  \frac{
                      q_t^2 \, [1- q_t^2/(4p_{{\rm F}e}^2)]}
                     {(q_r - \Omega)
                    (q_r - \Omega - {\bf q}_t \cdot {\bf k}/p_{{\rm F}e})}~,
\label{Lep1}
\end{eqnarray}
where we have introduced integration over
d${\bf q} = {\rm d} \varphi_{q} \, {\rm d} q_t \, q_t \,
{\rm d} q_r$ with cylindrical axis along   ${\bf p}_1$ and
integration over d${\bf k} = {\rm d} \varphi  \,  k_t \,
{\rm d} k_t \, {\rm d} \omega$ with cylindrical axis along ${\bf p}''_1$.
In the latter integration we have used the equality
d$k_r = {\rm d} \omega$;
$\varphi $ is an azimuthal angle of ${\bf k}$
with respect to the ${\bf p}''_1$-axis.

Consider the case of
$q_s \ll p_{{\rm F}e}$ typical for
the NS cores, where $q_s$ is
the plasma screening momentum (Sect.\ 2).
Integration in Eq.~(\ref{Lep1})
can be simplified because the
main contribution comes from the values
$q_s \la q \ll p_{{\rm F}e}$ which correspond to the
{\em small-angle approximation}.
In this
approximation,
we can omit $q_t^2/(2p_{{\rm F}e})^2$ in the
numerator and ${\bf q}_t \cdot {\bf k}/p_{{\rm F}e}$ in the denominator
of the integrand.
Furthermore we may set
$1/(\varepsilon_1 \varepsilon'_1)=1/p_{{\rm F}e}^2 \ $
in the integrand of (\ref{Lep1})
and take into account that
the energy transfer $\Omega$
from an electron to a proton is
\begin{equation}
     \Omega =  \varepsilon_2 - \varepsilon'_2
     \approx  \,
    \frac{1}{m_p^\ast }
    \left( {\bf p}_2 \cdot {\bf q} + \frac{q^2}{2} \right) \,
    \approx \, \frac{p_{{\rm F}e} q }{m_p^\ast}
    \cos \vartheta_{q2}~,
\label{Omega}
\end{equation}
where $\vartheta_{q2}$ is an angle between
${\bf q}$ and ${\bf p}_2$.
Then the integrand in Eq.~(\ref{Lep1})
depends only on the relative
positions of  ${\bf p}_2$ and ${\bf q}$.

First we can fix positions of ${\bf p}_1$,
${\bf p}_2$ and ${\bf q}$ and perform the following integrations:
\begin{equation}
         \int_0^{2\pi} {\rm d} \varphi \,
         \int_0^{k_0} {\rm d} k_t \, k_t (k_0^2 - k_t^2) \, = \,
        2 \pi (q_r - \Omega)^2 \left[ \omega - \frac{1}{2}
        (q_r - \Omega) \right]^2~ .
\label{Int-kt}
\end{equation}
A sequence of
angular integrations yields
$\int$  d$\Omega_1 \int$
d$\varphi_q \, \int$
d$\Omega_2 \, \ldots =$ $ 16\pi^3 \, \int$
d$\cos \vartheta_{q2} \, \ldots$
Finally, we are left with
the only
angular integration
in Eq.~(\ref{Lep1})
, that 
over  d$\cos \vartheta_{q2}$.
According to Eq.~(\ref{Omega}),
the variation range of
$\cos \vartheta_{q2}$ is related to
variation range of $\Omega$.
Let us consider two extreme cases.\\

{\em Low temperature limit}
$(T \ll q_s p_{{\rm F}e}/m_p^\ast \ll q_s)$ corresponds to
thermal energies $T$ much less
than recoil energy of protons.
This case is most
important for practical use [15].
Let us replace integration over $\cos \vartheta_{q2}$
by integration over $\Omega$:
$\cos \vartheta_{q2} = m_p^\ast \Omega/(q_t p_{{\rm F}e})$,
where $(q_r -2\omega) \leq  \Omega \leq q_r$.
The Pauli principle
strongly
restricts possible values of $\Omega$ as well as
final states of electrons and protons.

The condition $T \ll q_s$ leads to the inequalities
$q_r \ll q_t$, $\omega \ll q_t$, $\Omega \ll q_t$.
The integration over
d$q_t$ in Eq.~(\ref{Lep1})
yields
\begin{equation}
              \int_0^{2p_{{\rm F}e}} \, {\rm d} q_t \,
              \frac{q_t^2}{(q_t^2 + q_s^2)^2} \,
              \simeq \, \frac{1}{2p_{{\rm F}e}}
              \left( \frac{\pi}{4 y_s} -1 \right)~,
\label{Int-qt}
\end{equation}
where $y_s$ is given by Eq.\ (\ref{y_s}).

The integrations over $\varepsilon_1$ and $\varepsilon_2$
are standard:
\begin{eqnarray}
      &&    \int_{-\infty}^{+\infty} \; {\rm d} \varepsilon_1 \;
            \int_{-\infty}^{+\infty} \; {\rm d} \varepsilon_2 \;
            f_1 \, f_2 \, (1 - f'_1) \, (1 - f'_2)
\nonumber \\
      &&  =  \frac{(\omega + \Omega)}
             {{\rm e}^{(\omega + \Omega)/T} - 1}
              \; \frac{\Omega}
               {1 - {\rm e}^{- \Omega/T}} \quad
               {\rm for} \, \, \, \Omega>0~, \,  (\omega + \Omega)>0;
\nonumber    \\
      && =  \frac{(|\Omega|- \omega )}
                   {1 - {\rm e}^{-(|\Omega|-\omega )/T}}
                      \; \frac{|\Omega|}
               {{\rm e}^{|\Omega|/T} - 1} \quad
               {\rm for} \, \,\, \Omega< 0, \, (\omega + \Omega)< 0,
\label{IntFermi}
\end{eqnarray}
%
where $\varepsilon'_1 = \varepsilon_1 - \omega - \Omega$ and
$\varepsilon'_2 = \varepsilon_2 + \Omega$.
Now we are left with integrations over
d$\omega$  d$\Omega$ d$q_r$
in six domains
restricted by the inequalities
$q_r \geq \Omega$ and
$\omega \geq (q_r - \Omega)/2 \geq 0 $ (see Eqs.~(\ref{OK})).
Three of them correspond to $q_r>0 \,\, $
$[ \, (a) \,\, \Omega \geq 0; \,\, (b) \,\,
\Omega < 0, \, \omega +\Omega \geq 0;
\,\, (c) \,\, \omega + \Omega <0 \, ]$,
while three others correspond to
$0 \geq q_r \geq \Omega \,\, $
$[ \, (d) \,\, \omega + \Omega \geq 0; \,\,
(e) \,\, 2\omega + \Omega \geq 0 > \omega + \Omega; \,\,
(f) \,\, 0> 2\omega + \Omega \geq q_r \, ]$.
Following Haensel et al.\ [24]  we can introduce
dimensionless variables
$u=q_r /T, \, w=\Omega/T, \, v=\omega/T$,
$v_0=(q_r-\Omega)/(2T)$.
Six domains of integration over d$v$ d$w$ d$v_0$
in Eq.\ (\ref{Lep1})
are reduced to four ones,
and integration over d$v_0$
is trivial $ \,\, (\int_0^v {\rm d} v_0 \, (v - v_0)^2 = v^3/3)$.

Finally, using Eqs.~(\ref{Lep1}), (\ref{Int-kt}),
(\ref{Int-qt}), and (\ref{IntFermi})
we obtain
\begin{equation}
                 L_{ep} \, \approx \, \frac{189 m_p^{\ast2} T^2}
                 {2^7 \pi^5 p_{{\rm F}e}^4 y_s} \, \eta \, = \,
                 {3 \pi^3 \over 40 y_s} \left( {m_p^\ast T \over
                 p_{{\rm F}e}^2}  \right)^2 \xi~,
\label{Lep2}
\end{equation}
where
\begin{eqnarray}
                \eta & = &
                \int_0^\infty {\rm d}v \,v^4 \left\{
                \int_0^\infty {\rm d} w \, \frac{ w \,(v+w)}
                {({\rm e}^{v+w} -1)(1- {\rm e}^{-w})}  \right.
\nonumber   \\
                      & + &
                 \int_0^v {\rm d} w \, \frac{ w \, (v+w)}
                 {({\rm e}^{v+w} -1)(1- {\rm e}^{-w})} \, + \,
                 \int_0^v {\rm d} w \, \frac{ w \, (v-w)}
                 {({\rm e}^{v-w} -1)({\rm e}^w - 1)} 
\nonumber  \\
                       & + &
             \left.
               \int_{2v}^\infty {\rm d} w \, \frac{ w \, (v-w)}
                 {(1 - {\rm e}^{v-w})({\rm e}^w - 1)}  \right\}
              \approx 1646.7~,
\label{eta}
\end{eqnarray}
as obtained by numerical integration, and
$\xi = 315 \, \eta /(16\pi^8) \approx 3.417$.

Kaminker et al.\ [15] proposed to estimate
$L_{ep}$ by rescaling the Coulomb logarithm $L_{\rm br}$
calculated by Haensel
et al.\  [24]  for scattering of
ultrarelativistic degenerate
electrons by
atomic nuclei in liquid state.
In the context of
$ep$ scattering, $L_{\rm br}$ is valid
at $T_{{\rm F}p} \ll  T  \ll T_{{\rm F}e} $.
The rescaling procedure
has been based on the similarity criterion:
$L_{ep} = L_{\rm br} \, \nu_{ep}/ \nu^{(0)}_{ep}$,
where $\nu_{ep}$ is an effective $ep$ collision
frequency 
for the conditions under consideration 
($T \ll q_s \la T_{{\rm F}p} \ll T_{{\rm F}e}$) and
$\nu^{(0)}_{ep}$ is an effective frequency of elastic
$ep$ collisions at $T \gg q_s$
(e.g.,\ [25]). For $T \ll q_s$,
the authors used 
the collision frequency $\nu_{ep}$ that
determines the electron thermal conductivity
(e.g.,\ [20]).
One can easily verify that the result
given by Eqs.~(\ref{Lep2}) and (\ref{eta}) differs
from the crude estimate by Kaminker et al.\ [15] by
a small factor $y_s^2$ given by (\ref{y_s}).
Thus Kaminker et al.\ [15] have
strongly (typically, by two orders of magnitude)
overestimated $Q_{ep}$, 
obtained in the present paper by
an accurate evaluation.
The nature of this overestimation comes from
using the effective thermal-conduction collision
frequency $\nu_{ep}$ in the rescaling procedure.
The thermal conduction at $T \ll q_s$ is associated
with small momentum transfers ($q \sim T$). Corresponding
collision frequency appears to be
greatly enhanced as compared to the frequency of collisions
with higher momentum transfers ($q \sim q_s$) which should have
been used in rescaling. For instance, it would be
appropriate to use the $ep$ collision frequency
which determines electron electric conductivity
(e.g.,\ [26]). Therefore, the rescaling
proposed in ref.\ [15] is basically
correct if one uses physically justified effective
collision frequencies.

Finally, we obtain (in the standard physical units)
\begin{eqnarray}
     Q_{ep} & = & { \pi^4 \xi \, G_{\rm F}^2 e^4 C_{e+}^2  m_p^{\ast2}
                  \over
           945 \, \hbar^9 c^8 \, y_s \, p_{{\rm F}e}^4 } \, n_p \,
           (k_{\rm B} T)^8  R_{ep}
\nonumber \\
    & \approx &  {3.69 \times 10^{14} \over y_s}
          \left({ m_p^\ast \over m_p } \right)^2
          \left( { n_0 \over n_p} \right)^{1/3} \,
          R_{ep} \, T_9^8~~{\rm erg~cm^{-3}~s^{-1}}~.
\label{Qep}
\end{eqnarray}
Here we have introduced the factor
$R_{ep}$ that describes suppression of the emissivity
by the proton superfluidity. 
We introduced also $T_9\equiv 
T/10^9~{\rm K}$.

If protons are normal ($T \geq T_{{\rm c}p}$), one has
$R_{ep}=1$ in Eq.\ (\ref{Qep}).
The factor $R_{ep}$ reduces
neutrino emission at $T < T_{{\rm c}p}$. It
should be the same as the factor $R_{np}$
that describes reduction of the
bremsstrahlung neutrino emission in $np$ collisions
by proton superfluidity [3]:
\begin{eqnarray}
     R_{ep} &  =  & {1 \over 2.732} \left[
       A \exp \left( 1.306 - \sqrt{(1.306)^2+y^2 } \right) \right.
\nonumber \\
       & + & \left.
       1.732 \,
       B^7 \exp \left( 3.303 - \sqrt{(3.303)^2+4 y^2 } \right) \right],
\nonumber   \\
          A & = & 0.9982 + \sqrt{(0.0018)^2+(0.3815y)^2}~,
\nonumber   \\
          B & = & 0.3949+ \sqrt{(0.6051)^2+(0.2666y)^2}~,
\label{Rep}
\end{eqnarray}
where  $y=\Delta /T$ (Sect.\ 2).

{\em Moderate temperature limit}
($q_s p_{{\rm F}e}/m_p^\ast \ll T \ll q_s$).
In this case thermal energy is much higher than the recoil
energy of protons. Therefore, as follows from Eq.~(\ref{Omega}),
$T \gg \Omega$. To estimate $L_{ep}$ in this limit we
can omit $\Omega$ in comparison with all quantities $\sim T$
in the integrand and in the integration limits
in Eq.~(\ref{Lep1}). Note that the condition
$k_0^2 \approx q_r (2\omega - q_r) \geq 0$
obviously gives $q_r \geq 0$
and $\omega \geq q_r/2$.
Then we easily perform
integrations in Eq.~(\ref{Int-kt}) and subsequent
integrations over d$\Omega_2$  d$\varphi_q$  and d$\Omega_1$.
In the limit of $\Omega \to 0$
standard integration yields
\begin{equation}
        \int {\rm d} \varepsilon_1 \int {\rm d} \varepsilon_2
        \,\,   f_1 \, f_2 \,(1- f'_1) \, (1 - f'_2)
        \, = \, \frac{\omega \, T}{{\rm e}^{\omega/T} - 1}~.
\label{Int-eps}
\end{equation}
which gives
\begin{eqnarray}
                  L_{ep} & = &
                  \frac{567 m_p^\ast \, T }{2^4 \pi^6 p_{{\rm F}e}^2}
                  \, \, \int_0^{2p_{{\rm F}e}} {\rm d} q_t \,
                  \, \frac{q_t^3}{(q_t^2 + q_s^2)^2} \,
                  \int_0^\infty {\rm d} u \,
                  \int_{u/2}^\infty {\rm d} v \,
                  \frac{v^2 (v - u/2)^2}{{\rm e}^v - 1}
\nonumber   \\
                   & = &  \,\,
                   3 \left( \frac{m_p^\ast T}{p_{{\rm F}e}^2}
                   \right) \, L^{(0)}_{ep},
\nonumber  \\
                 \, & \, \, & \,
\nonumber   \\
              L_{ep}^{(0)} & = & \int_0^{2p_{{\rm F}e}} {\rm d} q_t
                   \, \frac{q_t^3}{(q_t^2 + q_s^2)^2}
                   \,\, \approx \,\,
                   \ln \left(\frac{1}{y_s} \right) - \frac{1}{2}~.
\label{Lep3}
\end{eqnarray}
Here $L_{ep}^{(0)}$ is close to
the Coulomb logarithm for
neutrino
bremsstrahlung due to collisions of electrons with
nondegenerate protons in the low--temperature limit
($ T_{{\rm F}p} \la T \ll q_s \ll T_{{\rm F}e} $).
The factor $3 m_p^\ast T /p_{{\rm F}p} ^2$
in Eq.\ (\ref{Lep3}) appears due to degeneracy of protons
in the case considered here.
As a result, we obtain the emissivity
(in the standard physical units)
\begin{eqnarray}
     Q_{ep} & = & {8 \pi G_{\rm F}^2 e^4 C_{e+}^2  m_p^\ast \over
           189 \hbar^9 c^8 \, p_{{\rm F}e}^2 } \, n_p \,
           (k_{\rm B} T)^7 \, L_{ep}^{(0)} \, R_{ep}
\nonumber   \\
          \, & \, \, & \,
\nonumber \\
          & \approx &  1.89 \times 10^{17}
          \left({ m_p^\ast \over m_p } \right)
          \left( { n_p \over n_0} \right)^{1/3} \,
          L_{ep}^{(0)} \, R_{ep} \,  T_9^7~~{\rm erg~cm^{-3}~s^{-1}}.
\label{Qep2}
\end{eqnarray}
Note that at {\em low temperatures}
$Q_{ep}$ has the temperature dependence
$Q_{ep} \propto T^8$, which differs from the case of
{\em moderate temperatures},
$Q_{ep} \propto T^7$, 
as well as from the case of nondegenerate protons,
$Q_{ep} \propto T^6$ ($T \ga T_{{\rm F}p} $).

The case $ q_s  \ll T \la  T_{{\rm F}p} $ has no importance
for astrophysical applications.
It can be shown that in this case
the emissivity is again given by Eq.~(\ref{Qep2})
with somewhat different (temperature dependent)
Coulomb logarithm $L_{ep}^{(0)}$.

\section{Neutrino $ee$ bremsstrahlung}
Let us consider now 
the neutrino emissivity
$Q_{ee}$ for $ee$ bremsstrahlung.
We restrict ourselves 
 by the same condition
$q_s \ll p_{{\rm F}e}$
(small--angle approximation)
as for $ep$ bremsstrahlung.
In contrast to the previous case,
only the region of relatively small temperatures,
$T \ll q_s$, is important for practical applications.

The integrand in Eq.\ (\ref{Lee})
depends
on the relative
positions of ${\bf p}_1$, ${\bf p}_2$, ${\bf q}$ and  ${\bf k}$.
In particular it depends on angle $\Theta$ between ${\bf p}_1$
and ${\bf p}_2$. Thus we may direct ${\bf p}_1$ along the $z$-axis
and place ${\bf p}_2$ in the $(xz)$-plane.  
Let $(\vartheta_{q1},
\varphi_q)$  be polar angles of ${\bf q}$ and
$(\vartheta_{k1}, \varphi_k)$ be polar angles of ${\bf k}$.
Let us introduce also an angle $\vartheta_{q2}$ between
${\bf q}$ and ${\bf p}_2$, an angle $\vartheta_{k2}$
between ${\bf k}$ and ${\bf p}_2$, and 
a relative azimuthal angle $\varphi_{kq}$
between ${\bf k}$ and ${\bf q}$.

We can use  relations
(\ref{OK})
as for $ep$ bremsstrahlung.
While evaluating the
right hand side of
Eq.~(\ref{Lee})
it is useful to introduce
also an additional auxiliary integral
$\int {\rm d} \Omega \, \delta (\varepsilon_2-\varepsilon'_2+\Omega)$.
Using the relation $\varepsilon'_2 = |{\bf p}_2 +{\bf q}|
\approx p_2 (1 + (q/p_{{\rm F}e}) \, \cos \vartheta_{q2} )$ we can
present the $\delta$--function under this integral in
the form
\begin{equation}
           \delta (\varepsilon_2 - \varepsilon'_2 + \Omega)
          \approx \frac{1}{q} \, \delta  \left( \cos \vartheta_{q2} -
                  \frac{\Omega}{q} \right).
\label{delta}
\end{equation}
Equations (\ref{Lee}), (\ref{JXYZ}), (\ref{X}), (\ref{OK}),
and
%
%
(\ref{delta})
yield
\begin{eqnarray}
               L_{ee} & = & L_{ee}^{(1)} - L_{ee}^{(2)}  \approx
                   \frac{567}{2^9
                      \pi^{10} T^6 p_{{\rm F}e}}
                \int {\rm d} \varepsilon_1
                \int {\rm d} \varepsilon_2
                \int  {\rm d} \Omega_1
                \int {\rm d} \Omega_2
                \int_0^{2\pi} {\rm d} \varphi_q
\nonumber    \\
                \, & \,\, & \,
\nonumber     \\
                 \,  & \times &  \,
               \int_0^{2 p_{{\rm F}e}} {\rm d} q_t \, q_t
               \int_0^\infty {\rm d} q_r
               \int_{-\infty}^{q_r} {\rm d} \Omega
               \int_0^{2\pi} {\rm d} \varphi
               \int_{(q_r - \Omega)/2}^\infty {\rm d} \omega
                \int_0^{k_0} {\rm d} k_t \, k_t
 \nonumber   \\
                    \, & \, & \,
\nonumber    \\              
                       & \times  &
                   \frac{1}{q} \, \, \delta \left( \cos \vartheta_{q2}
               \,  -  \, \frac{\Omega}{q} \right) \,
               \frac{( 1 \, - \, \cos \Theta )^2}
               { ( q_t^2 + q_r^2
                    - \Omega^2 + q_s^2 )^2} \,
                  \frac{(k_0^2 - k_t^2) q_t^2 }
                     {(q_r - \Omega)^2}
\nonumber   \\
                         \, &  \,  &  \,   
\nonumber   \\             
                       & \times &         
                  \left[ 1 - \frac{1+ (C_{e-}/C_{e+})^2}{2} \,
                  \frac{(q_r - \Omega)}
                  {\omega - k \cos \vartheta_{k2}} \right]
                  \omega f_1 f_2 (1-f'_1)(1-f'_2)~.
\label{Lee1}
\end{eqnarray}
Here we have introduced the same notations as in Eq.~(\ref{Lep1}),
and we have divided $L_{ee}$ into two parts,
$L_{ee}^{(1)}$ and $L_{ee}^{(2)}$.
The first part,
$L_{ee}^{(1)}$, corresponds to the first term (equal to 1)
in square brackets which comes from
$X$ in Eq.~(\ref{Lee}) (``direct" $ee$ scattering).
The second part,
$L_{ee}^{(2)}$, corresponds to
the second term
which comes from $I_1$ in Eq.~(\ref{Lee})
(interference
of amplitudes with exchanging electrons 1 and 2).
It can be shown that
the terms $Y_1$ and $Z_1$ in Eq.~(\ref{Lee}) (interference
of amplitudes with mutually transposed electron final states)
contain small parameters  $\sim q_r/p_{{\rm F}e}$  and
$\sim (q_t/p_{{\rm F}e})^2$
and can be neglected.

Let us use the relationship
$\cos \vartheta_{q2} =
\cos \Theta \cos \vartheta_{q1} +
\sin \Theta \sin \vartheta_{q1} \cos \varphi_{q}$
in Eq.~(\ref{delta})
and express $\cos \vartheta_{q2}$
in the $\delta$-function
through integration variables 
$q_t$, $q_r$, $\varphi_q$, and $\Theta$ in Eq.~(\ref{Lee1}).
We also need to express $\cos \vartheta_{k2}$
through $k_r$, $k_t$, $\varphi$, and $\Theta$.
For this purpose we
take into account additional
approximate relations  (justified at $p_{{\rm F}e} \gg q_t$)
\begin{eqnarray}
       &&    \cos \vartheta_{q2}     \approx
            (q_r \, \cos \Theta \;  +
            q_t \sin \Theta \cos \varphi_q)/q~,
\nonumber   \\
       &&     \cos \vartheta_{k2}  \approx
            (k_r \, \cos \Theta +
            k_t \sin \Theta \cos \varphi_k)/k~,
\nonumber  \\
       &&    \cos \varphi_k    \approx
            \sin \varphi_{kq}
            \approx \sin \varphi~.
\label{angles}
\end{eqnarray}
The third relationship is valid
at  $q_r \ll q_s \la q_t$, $\Omega \ll q_t $,
and $\omega  \ll q_t$.
First we evaluate $L_{ee}^{(1)}$
in Eq.~(\ref{Lee1}).
The integration over d$\varphi$  is trivial,
the integration over $k_t$ is reduced
to (\ref{Int-kt}),
and the integration over d$\varphi_q$
can be done with help
of the delta-function 
\begin{equation}
          \delta \left( \cos \vartheta_{q2}  - 
                      \frac{\Omega}{q} \right) \, = \,
           \frac{q}{q_t \, \sin \Theta} \,
           \delta \left( \cos \varphi_q -
           \frac{\Omega - q_r \cos \Theta}{q_t \, \sin \Theta}
           \right)~.
\label{delta-varphi}
\end{equation}
Now we can integrate over d$\Omega_1$ and d$\Omega_2$ in a
straightforward manner:
\begin{equation}
\int_{4\pi}\int_{4\pi} {\rm d} \Omega_1 {\rm d} \Omega_2
\, (1- \cos \Theta)^2 (\sin \Theta)^{-1} = 12 \pi^3.
\end{equation}
The integrations over
d$\varepsilon_1$ and d$\varepsilon_2$
can be performed using Eq.~(\ref{IntFermi}).
We are left with a 4-dimensional integral
(over d$\omega$ d$\Omega$ d$q_r$ d$q_t$).
Using Eq.~(\ref{Int-qt}) and introducing
the same dimensionless variables as in Eqs.~(\ref{Lep2}) and
(\ref{eta}), we obtain
\begin{equation}
                L_{ee}^{(1)}  \approx
                 \frac{567 \, T^2 }
                 {2^8 \pi^5 \, p_{{\rm F}e}^2 \, y_s} \,\, \eta
                 \, = \,
                 \frac{3 \pi^3}{16 \, y_s}
                 \left(  \frac{T}{p_{{\rm F}e}} \right)^2 \,
                 \xi_1~,
\label{Lee2}
\end{equation}
where $\eta$ is defined by Eq.~(\ref{eta}) and
$\xi_1 = 189 \eta /(16 \pi^8)= 2.05$.

Making use of
Eqs.~(\ref{angles}) and performing
straightforward but tedious calculations
we can also evaluate
$L_{ee}^{(2)}$. It appears to contain a small numerical factor
$\, \approx 0.05 \,$ as compared to $L_{ee}^{(1)}$.
Note that we have neglected
other interference terms in  Eq.~(\ref{Lee})
which could be of the same order of magnitude or even
larger than $L_{ee}^{(2)}$.
Therefore we will use only the main term (\ref{Lee2})
for applications.

Finally the neutrino emissivity $Q_{ee}$ can be written in the form
(in the standard units)
\begin{eqnarray}
     Q_{ee} & = & { \pi^4 \xi_1 G_{\rm F}^2 e^4 C_{e+}^2  \over
           378 \hbar^9 c^{10} y_s p_{{\rm F}e}^2 } \, n_e \,
           (k_{\rm B} T)^8
\nonumber \\
    & \approx &  {0.69 \times 10^{14} \over y_s}
          \left( { n_e \over n_0} \right)^{1/3} \,
          T_9^8~~{\rm erg~cm^{-3}~s^{-1}}.
\label{Qee}
\end{eqnarray}
As in the case of $Q_{ep}$, this result
differs from
a crude estimate given by Kaminker et al.\ [15]
by a small factor $y_s^2  \ll \, 1$ due to
an unfortunate use of
inappropriate
thermal-conduction $ee$ collision frequency in the rescaling
procedure.

Let us emphasize that the emissivity $Q_{ep}$
is reduced exponentially by strong proton
superfluidity, while $Q_{ee}$
is affected by the superfluidity
in a much weaker manner,
only through the plasma screening
parameter (\ref{y_s}).
If protons are normal, they provide the major contribution
into the plasma screening. If they are strongly superfluid,
a weaker electron screening becomes important,
which {\em enhances} $Q_{ee}$, but not
to a great extent (see below).

Let us remind that the neutrino emissivity $Q_{ee}$ for
nondegenerate electrons in the neutron
star crusts has been calculated
by Cazzola \& Sagion [16, 17].
We have compared their numerical results
(Fig.\ 2 in ref.\ [17]) with
our  Eq.\ (\ref{Qee})
for several temperatures ($T_9 = 5, 7, 8, 10$)
at those values of
$\rho \ga 10^8$ g cm$^{-3}$, where the relativistic electrons
start to become degenerate
and both approaches are expected
to be qualitatively valid.
We have obtained reasonable agreement
of the results.

Finally let us make a few remarks about neutrino bremsstrahlung
reactions involving muons, $\mu^-$.
Muons can
appear in sufficiently dense matter of neutron star cores,
provided the electron number density exceeds the threshold value
$n_e > (m_\mu c)^3/(3 \pi^2 \hbar^3) = 0.0052$ fm$^{-3}$,
where $m_\mu$ is the muon mass. For realistic equations
of state, this happens at $\rho \ga$ (1.5--2)$ \, \rho_0$.
The muons, like electrons, form almost ideal degenerate
gas, but they are mainly non-relativistic
(near the threshold density) or mildly relativistic at
much higher densities.
Beta-equilibrium condition
implies $\mu_\mu = \mu_e$, where $\mu_\mu$ is the
muon chemical potential.
The number density of muons is much smaller than
the number density of neutrons or electrons for all densities.
In muonic matter, several new
neutrino bremsstrahlung reactions are allowed,
resulting from
$\mu e$, $\mu \mu$, $\mu p$ scattering. The neutrino emissivities
in these reactions can be evaluated in the same manner as
the emissivities due to $ep$ and $ee$ neutrino bremsstrahlung.

For instance, in the nonrelativistic limit
($p_{\rm F\mu} \ll m_\mu c$) at $T \ll q_s$,
in analogy with Eq.\ (\ref{Qep}),
we obtain
\begin{eqnarray}
         Q_{e\mu} & = & \frac{\pi^4 \xi \, G_{\rm F}^2 e^4 C_{e+}^2
         m_{\mu}^2}{945 \hbar^9 c^8 y_s p_{\rm F\mu}^3
         p_{{\rm F}e}} \, n_{\mu} (k_{\rm B}T)^8
\nonumber     \\
         & \approx &
         \frac{4.68 \times 10^{12}}{y_s} \left(
         \frac{n_0}{n_e} \right)^{1/3}
         T_9^8~~~{\rm erg~cm^{-3}~s^{-1}}~,
\label{Qemu}
\end{eqnarray}
where $n_\mu$ is the muon number density.
Equation (\ref{Qemu}) differs from Eq.\ (\ref{Qep}) by a small
factor $(m_\mu / m_p^\ast)^2$.

One can see that, under considered conditions,
the neutrino
emissivity $Q_{e\mu}$ is much smaller
than the emissivity
$Q_{ee}$  due to $ee$ scattering
(cf. Eqs.\ (\ref{Qemu}) and (\ref{Qee})).

Our estimates show that the neutrino bremsstrahlung reactions
involving muons are less efficient than
analogous reactions involving electrons,
for all densities
typical for neutron star cores.

\section{Discussion}
Let us discuss the efficiency of various neutrino
emission mechanisms in superfluid NS cores. For illustration,
we adopt
a moderately stiff equation of state
proposed by Prakash et al.\ [27]
(the version with compression modulus $K_0 = 180$ MeV and with
the same simplified 
symmetry energy 
$S_V$ which
was used by Page \& Applegate [28]).
We assume that 
dense neutron-star matter
 consists of
neutrons, protons and electrons (no muons and hyperons).
The nucleon effective masses will be set equal to $0.7$
of the masses of bare particles.
The equation of state allows the direct Urca process
to operate at $\rho \geq 4.64 \, \rho_0$.
We consider two densities, $\rho=2 \, \rho_0$,
as an example of the standard neutrino
cooling (the direct Urca is forbidden),
and  $\rho= 5 \, \rho_0$,
representing the case of
the cooling enhanced by the direct Urca process.
Both nucleon species, neutrons and protons,
are supposed to form superfluids.
We consider a triplet-state neutron pairing
($^3$P$_2$), with zero projection of the total momentum ($m_J=0$)
of a Cooper pair onto quantization axis,
and a singlet-state proton pairing ($^1$S$_0$).
Since the critical temperatures $T_{{\rm c}n}$ and $T_{{\rm c}p}$
of both superfluids, as calculated from microscopic theories,
are very model dependent
(e.g., refs.\ [5, 6, 29-32]),
we do not make any specific choice of
microscopic superfluid model but treat
$T_{{\rm c}n}$ and $T_{{\rm c}p}$ as free parameters.

Figures 1 and 2
show temperature dependence of the neutrino emissivities
for various neutrino production mechanisms
at $ \, \rho = 2\rho_0 \,$ and $ \,\rho = 5\rho_0 \, $,
respectively,
assuming rather strong
neutron
and moderate
proton superfluids,
$ \,T_{{\rm c}n}=8 \times 10^9 \,$~K  and
$\, T_{{\rm c}p}=2.5 \times 10^9 \,$~K.
%
\begin{figure}[ht]
\begin{center}
\leavevmode
\epsfysize=100mm
\epsfbox{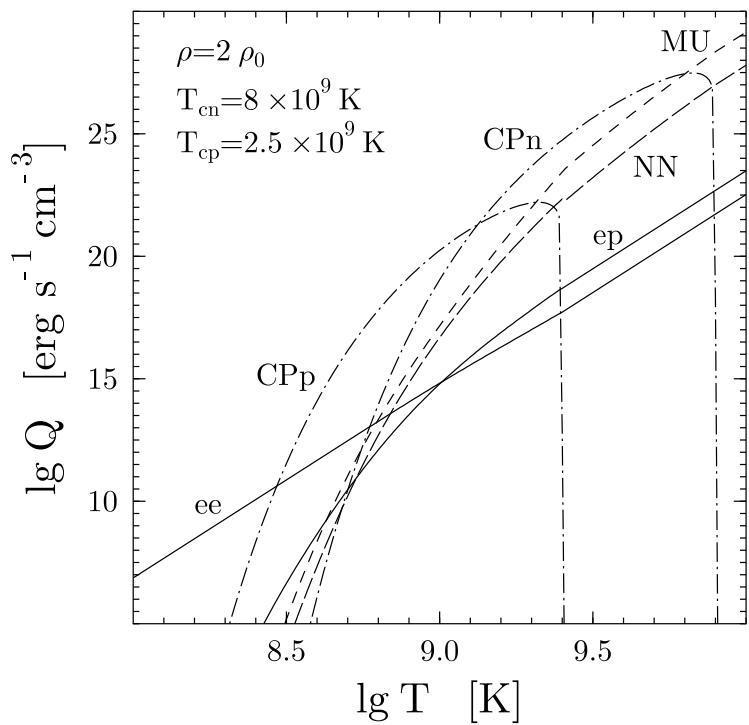}
\end{center}
\caption[]{
     Temperature dependence of
     neutrino energy emission rates for various neutrino reactions
     in $npe$ matter at $\rho = 2 \, \rho_0$,
     $T_{{\rm c}n}= 8 \times 10^9$~K and
     $T_{{\rm c}p}= 2.5 \times 10^9$~K.
     The neutrino reactions are: modified Urca
     (MU, sum of $n$ and $p$ branches),
     nucleon-nucleon bremsstrahlung (NN, sum of
     $nn$, $np$ and $pp$ branches), $ee$ and $ep$ bremsstrahlung,
     Cooper pairing of neutrons (CPn) and protons (CPp).
     }
\label{fig1}
\end{figure}
%

In Fig.\ 1 we show neutrino emission from the modified Urca
process (\ref{Mur})
(sum of the proton and neutron branches),
nucleon-nucleon bremsstrahlung (\ref{NN})
(sum of $nn$, $np$ and $pp$ reactions),
Cooper pairing of neutrons and protons (\ref{CP}),
and $ep$  and $ee$  bremsstrahlung  processes (\ref{ep}) and (\ref{ee}).
The emissivities in the last two reactions
are obtained in Sects.\ 4 and 5.
The emissivities of the modified Urca and nucleon-nucleon
bremsstrahlung reactions (\ref{Mur}) and (\ref{NN})
are taken as from 
Levenfish \& Yakovlev [7],
with proper account for suppression of the reactions
by the 
neutron and proton superfluidities
[3].
The neutrino energy emission rate due to pairing of neutrons
is plotted in accordance with the results
by Yakovlev et al.\ [11, 12].
The neutrino emission due to pairing
of protons is calculated
taking into  account the relativistic corrections
[13]  to
the non-relativistic result [11, 12].
The allowance for the relativistic corrections
increases production of neutrinos due to proton pairing
by more than one order of magnitude.
The neutrino emissivity due to $ep$ bremsstrahlung
is calculated in the {\em low temperature limit}
(see Sect.\ 4) from Eqs.\ (\ref{Qep}) and (\ref{Rep}).
We use Eq.\ (\ref{Qee}) to obtain the neutrino emission
due to $ee$ bremsstrahlung.

At $ T > T_{{\rm c}n}$ 
the neutron-star matter is  nonsuperfluid,
so that
the modified Urca process and nucleon-nucleon bremsstrahlung
play leading roles.
It can be seen 
that
the nucleon superfluidity
reduces the Urca and bremsstrahlung
neutrino reactions involving nucleons
((\ref{Mur}), (\ref{NN}), and
(\ref{ep})) at $T \la 10^9$~K.
On the other hand, the
nucleon superfluidity produces
two bumps of the partial
neutrino emissivities
at temperatures slightly below two critical temperatures,
$T_{{\rm c}n}$ and $T_{{\rm c}p}$,
due to Cooper pairing of neutrons and protons,
respectively.
The emission due to pairing of neutrons dominates
over other mechanisms at $T \la 6 \times 10^9$~K,
while the emission due to pairing of protons
dominates at $\, T \la 1.5 \times 10^9$~K.
If the Cooper pairing emission were absent,
the total
neutrino emissivity at $T \la 10^9$~K would be
2--4 orders of magnitude smaller due to
strong suppression by the nucleon superfluidity.
The Cooper pairing can easily
turn suppression into enhancement
at not too low temperatures [11].
However, the Cooper-pairing
neutrino reactions are
eventually suppressed
 by superfluidity,
when temperature falls down much below
the critical temperature
of neutrons or protons ($T \la 0.2 \, T_{{\rm c}N}$).
For $T \ga 10^9$~K,
the neutrino emission due to $ep$ bremsstrahlung
turns out to be more intense
than that due to $ee$
bremsstrahlung but both processes are much weaker
than the other neutrino processes.
With decreasing $T$, the situation becomes
drastically different.
At $T \la 2.5 \times 10^8$~K,
$ee$ bremsstrahlung
becomes the leading mechanism
of neutrino emission
since it is the only mechanism which is almost
insensitive to the superfluid state of matter and suffers
no exponential suppression. 
Let us emphasize that
its role is even more important for stronger superfluidities
(for higher $T_{{\rm c}n}$ and $T_{{\rm c}p}$).

%
\begin{figure}[ht]
\begin{center}
\leavevmode
\epsfysize=100mm
\epsfbox{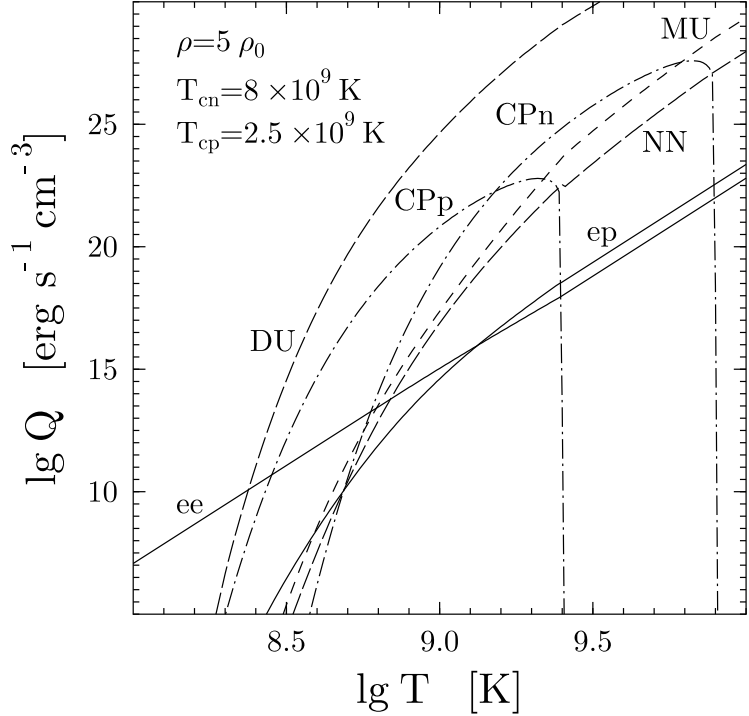}
\end{center}
\caption[]{
     Same as in Fig.\ 1 but
     for  $\rho = 5 \, \rho_0$.
     Direct Urca (DU) process is operative, in addition
     to neutrino reactions shown in Fig.\ 1.
     }
\label{fig2}
\end{figure}

In Fig.\ 2 we plot the contributions from the same
neutrino reactions
as in Fig.\ 1 and add the emissivity produced by the
powerful direct Urca process (\ref{Dur}).
The reaction (\ref{Dur}) is treated as described
by Levenfish \& Yakovlev [7], 
suppressed by nucleon superfluidity as in [21].
The direct Urca process is seen
to dominate all other reactions at
$T \ga 2 \times 10^8$~K. However, in spite of its high
efficiency in nonsuperfluid matter,
it is strongly suppressed
at low $T$ by nucleon superfluidity,
so that $ee$ bremsstrahlung
becomes the leading neutrino emission mechanism at
$T \la 2 \times 10^8$~K.
Therefore
$ee$ bremsstrahlung can play leading role even in the case
of neutrino cooling in which the direct Urca process is
 open (by momentum conservation rules)
but
strongly reduced
by superfluidity.
Since the superfluidity
reduces the direct Urca emissivity,
it reduces also strong difference
between the standard and enhanced cooling.
The temperature
range of the $ee$  bremsstrahlung  
domination  is 
narrower for 
rapid cooling.

%
\begin{figure}[ht]
\begin{center}
\leavevmode
\epsfysize=100mm
\epsfbox{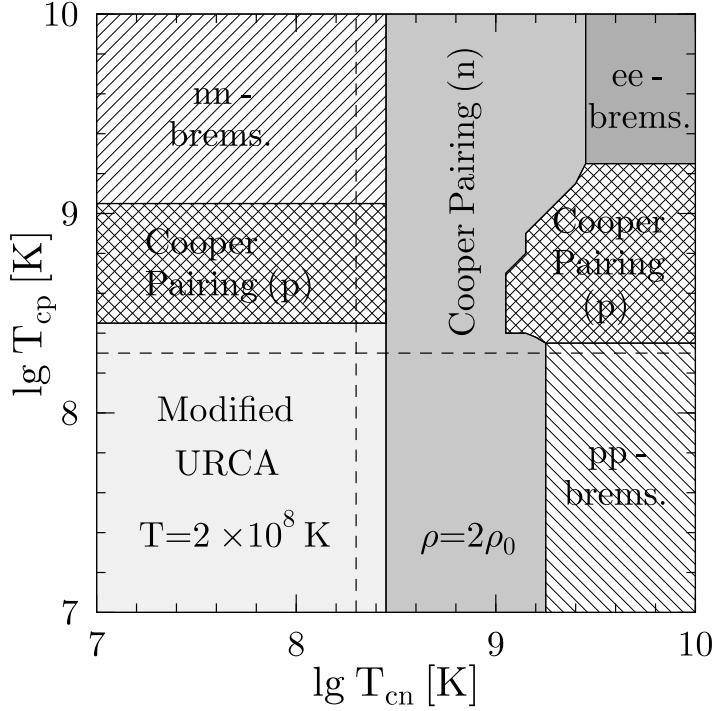}
\end{center}
\caption[]{
     Domains of the neutron and proton critical temperatures
     $T_{{\rm c}n}$ and $T_{{\rm c}p}$ in which different neutrino emission
     mechanisms dominate in a NS core at
     $\rho = 2 \, \rho_0$ and $T= 2 \times 10^8$~K.
     Dashes show the lines of $T=T_{{\rm c}n}$ and $T=T_{{\rm c}p}$.
     }
\label{fig3}
\end{figure}
\begin{figure}[ht]
\begin{center}
\leavevmode
\epsfysize=100mm
\epsfbox{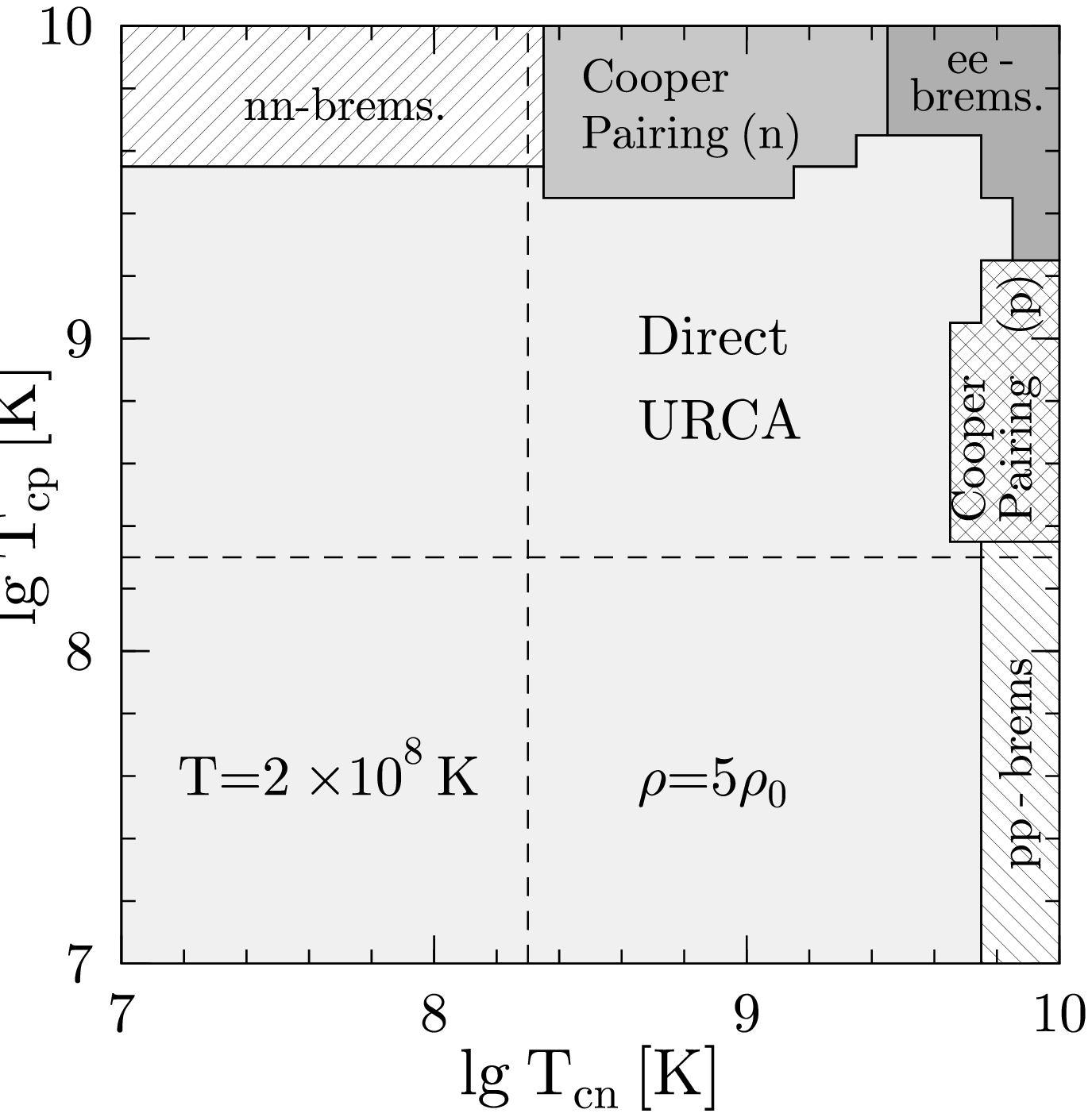}
\end{center}
\caption[]{
     Same as in Fig.\ 3 but for the neutrino
     emission enhanced by the direct Urca process
     at $\rho = 5 \, \rho_0$.
     }
\label{fig4}
\end{figure}
%
Figures 3 and 4 display
the domains of $T_{{\rm c}n}$ and $T_{{\rm c}p}$,
where different neutrino production mechanisms dominate
at $T=2 \times 10^8$~K
for the standard and
rapid cooling, respectively.
Dashes show the lines $T=T_{{\rm c}n}$ and $T=T_{{\rm c}p}$.
In the left lower squares restricted by these lines,
neutron-star matter
is nonsuperfluid. Accordingly, the modified Urca process
in Fig.\ 3 and the direct Urca process in Fig.\ 4
are the most powerful neutrino reactions
in these domains.
In the regions to the right
of these  domains
neutrons are superfluid
but protons are not, while in the regions
above these  domains protons are
superfluid and neutrons are not.
In the upper right
squares enclosed
by the dashed lines both nucleon species,
$n$ and $p$, are superfluid.
One can see a variety of dominant mechanisms
regulated by superfluidity.
If both $n$ and $p$ superfluidities
are relatively strong (upper right corners of Figs.\ 3 and 4),
they switch off all the neutrino emission mechanisms involving
nucleons, and therefore,
the neutrino emission due to
$ee$ bremsstrahlung dominates.
This region
becomes wider with
decreasing temperature
 $T$ (e.g., during  NS
cooling).
Therefore, one of two reactions considered in this article,
neutrino $ee$ bremsstrahlung,
can be the dominant neutrino
reaction in 
both cases of standard and
enhanced cooling.
We will not
discuss all variety of mechanisms which may dominate at
different $T_{{\rm c}n}, T_{{\rm c}p}$ and $T$
but notice that all of them  are
easily implemented into the cooling codes since
their emissivities are described  by simple
fitting formulae [3, 7, 11, 12, 21].
The present article supplements this description with
new neutrino reactions.

\newpage
\section{Conclusions}
We have analyzed two neutrino emission mechanisms
in the cores of neutron stars: neutrino $ee$ and $ep$
bremsstrahlung.
We have obtained simple
expressions for calculating the neutrino energy emission
rates in these reactions (Sects.\ 4 and 5).
We have shown (Sect.\ 6) that the
neutrino $ee$ bremsstrahlung
can dominate in
superfluid cores of NSs
in some domains of  parameters
of neutron-star matter.
Therefore this reaction should be included in simulations of
neutron star cooling.

One may expect
that $ee$ bremsstrahlung will be the main neutrino production
mechanism in highly superfluid cores
($T_{{\rm c}n} \ga 3 \times 10^9$~K
and $T_{{\rm c}p} \ga 3 \times 10^9$~K)
of sufficiently old NSs (age $t \ga 10^5$ yr)
during the transition from the neutrino cooling stage
to the photon cooling stage.

{\bf Acknowledgements.}
The authors are 
very  grateful
to D.G.\ Yakovlev for
numerous stimulating discussions
and
for the assistance in the preparation
of the text of the present paper. 
They are also grateful  to Kseniya Levenfish for the
assistance in
preparation of Figs.\ 3 and 4.
One of the authors (ADK)
acknowledges excellent working
conditions and hospitality of N.\ Copernicus Astronomical
Center in Warsaw.
This work was supported in part by the
RBRF (grant No. 99-02-18099a), INTAS (grant No. 96-0542),
 and KBN
(grant 2 P03D 014 13).

\end{document}